\documentstyle[aps,psfig]{revtex}
\begin{document}\twocolumn
\title{A robust verification of the quantum nature of light}
\author{Matteo G. A. Paris\thanks{E-mail address: {\tt paris@unipv.it}}}
\address{Quantum Optics $\&$ Information Group \\ 
INFM and Dipartimento di Fisica 'A. Volta', Universit\'a di Pavia, via Bassi 6, 
I-27100 Pavia, Italy}\date{\today} \maketitle\begin{abstract}
We present a conditional experiment involving a parametric 
amplifier and an avalanche photodetector to generate highly 
nonclassical states of the radiation field. The nonclassicality 
is robust against amplifier gain, detector efficiency and 
dark counts. At the output all the generalized Wigner 
functions have negative values, and this is exploited in order 
to reveal the nonclassicality through quantum homodyne tomography. 
\end{abstract}\vspace{3pt}\par
Nonclassical states of light are relevant in many fields, which 
ranges from fundamental tests of quantum mechanics to applications 
in quantum communication and measurements \cite{rev1}. In the last two 
decades several schemes to generate different kinds of nonclassical light 
have been suggested, and some of them have been implemented \cite{rev2}. 
However, with the exception of squeezing, the generation and the detection of 
nonclassical light is experimentally challenging. The aim of this letter 
is to suggest a simple experimental scheme to verify the quantum nature of
light in a state as nonclassical as a Fock number state. 
The main feature of our experimental scheme is its robustness against the 
possible imperfections of the setup, such as finite amplifier gain, 
nonunit detector efficiency and the occurrence of dark counts. \par
Let us consider an active $\chi^{(2)}$ crystal operating as a 
nondegenerate parametric amplifier (NOPA, for details of the experimental
setup see the end of the paper). The NOPA, pumped at frequency $\omega_P 
=\omega_a + \omega _b$, couples two modes $a$ and $b$ (idler and signal 
modes) via the medium nonlinearity. In the rotating-wave approximation, 
the evolution operator of the NOPA under phase-matching conditions can 
be written as  
$U_\lambda = \exp{\left[\lambda \left(a^\dag b^\dag-ab\right)\right]}$ 
where the "gain" $\lambda$ is proportional to the interaction-time, 
the nonlinear susceptibility, and the pump intensity. 
For vacuum input, 
we have spontaneous parametric down-conversion and the output state 
is given by the twin-beam
\begin{eqnarray}
|\psi\rangle = \sqrt{1-\xi^2} \sum_{n=0}^\infty \xi^n |n\rangle_a 
\otimes |n\rangle_b \qquad \xi = \tanh\lambda
\label{twb}\;.
\end{eqnarray}
We now consider the situation in which one of the two beams (say, mode $b$) 
is revealed by an ideal avalanche {\sc on/off} photodetector, i.e. a detector 
which has no output when no photon is detected and a fixed output when one or more 
photons are detected. The action of an {\sc on/off} detector is described 
by the two-value POVM
\begin{eqnarray}
\hat\Pi_0\doteq\sum_{k=0}^\infty(1-\eta_{\sc a})^k|k\rangle\langle
k| \qquad\quad \hat\Pi_1\doteq \hat{{\bf I}} -\hat\Pi_0\label{pom}\;
\end{eqnarray}
$\eta_{\sc a}$ being the quantum efficiency. The outcome "1" 
(i.e registering a "click" corresponding to one or more incoming 
photons) occur with probability 
\begin{eqnarray}
P_1 = \langle\psi | \hat{{\bf I}} \otimes \hat \Pi_1 | \psi\rangle 
= \frac{\eta_{\sc a} \xi^2}{1-\xi^2 (1-\eta_{\sc a})} \label{probs}\;
\end{eqnarray}
and correspondingly, the conditional output states for 
the mode $a$ is given by
\begin{eqnarray}
\hat\varrho_1= \frac1{P_1} \hbox{Tr}_b \left[ |\psi\rangle\langle\psi | 
\: \hat{{\bf I}} \otimes \hat \Pi_1\right]
\label{cond}\;.
\end{eqnarray}
In the Fock basis we have
\begin{eqnarray}
\hat\varrho_1 = \frac{1-\xi^2}{P_1}
\sum_{k=1}^\infty \xi^{2k} \left[1-(1-\eta)^k\right]\: |k 
\rangle\langle k|\label{fock}\;.
\end{eqnarray}
The density matrix in Eq. (\ref{fock}) describes a {\em pseudo}-thermal 
state, where the vacuum component has been removed by the conditional 
measurement. Such a state is highly nonclassical, as also discussed in Ref. 
\cite{man}. In the limit of low gain $\lambda \ll 1$ the conditional 
state $\hat\varrho_1$ approaches the number state $|1\rangle\langle 1|$ with 
one photon. 
The Wigner function $W(\alpha)$ of the state (\ref{cond}) exhibits 
negative values for any gain $\lambda$ and 
quantum efficiency $\eta_{\sc a}$. In particular, in the origin of the phase
space we have
\begin{eqnarray}
W(0)=-\frac{2}{\pi}\:\frac{1-\xi^2}{1+\xi^2}\:\frac{1-\xi^2(1-\eta_{\sc a})}
{1+\xi^2(1-\eta_{\sc a})}\label{wig0}\;.
\end{eqnarray} 
One can see that also the generalized Wigner function for $s$-ordering
$W_s(\alpha) = -2/(\pi s)\int d^2 \gamma W_0(\gamma) \:  
\exp[2/s |\alpha -\gamma|^2]$ 
shows negative values for $s \in (-1,0)$. In particular 
one has 
\begin{eqnarray}
W_s(0) = &-& \frac{2(1+s)}{\pi}\frac{(1-\xi^2)}{(1-s)+\xi^2 (1+s)}\nonumber 
\\ &\times& \frac{1-\xi^2(1-\eta_{\sc a})}{(1-s)+\xi^2(1+s)(1-\eta_{\sc a})}
\label{ws0}\;.
\end{eqnarray}
If we take as a measure of nonclassicality the lowest index $s^\star$ 
for which $W_s$ is a well-behaved probability (regular, positive 
definite) \cite{ncl} Eq. (\ref{ws0}) says that for $\hat\varrho_1$ 
we have $s^\star=-1$, {\em i. e.} $\hat\varrho_1$ 
describes a state as nonclassical as a number state. 
\par
\begin{figure}[h]\hspace{-10pt}\begin{tabular}{lr}
\psfig{file=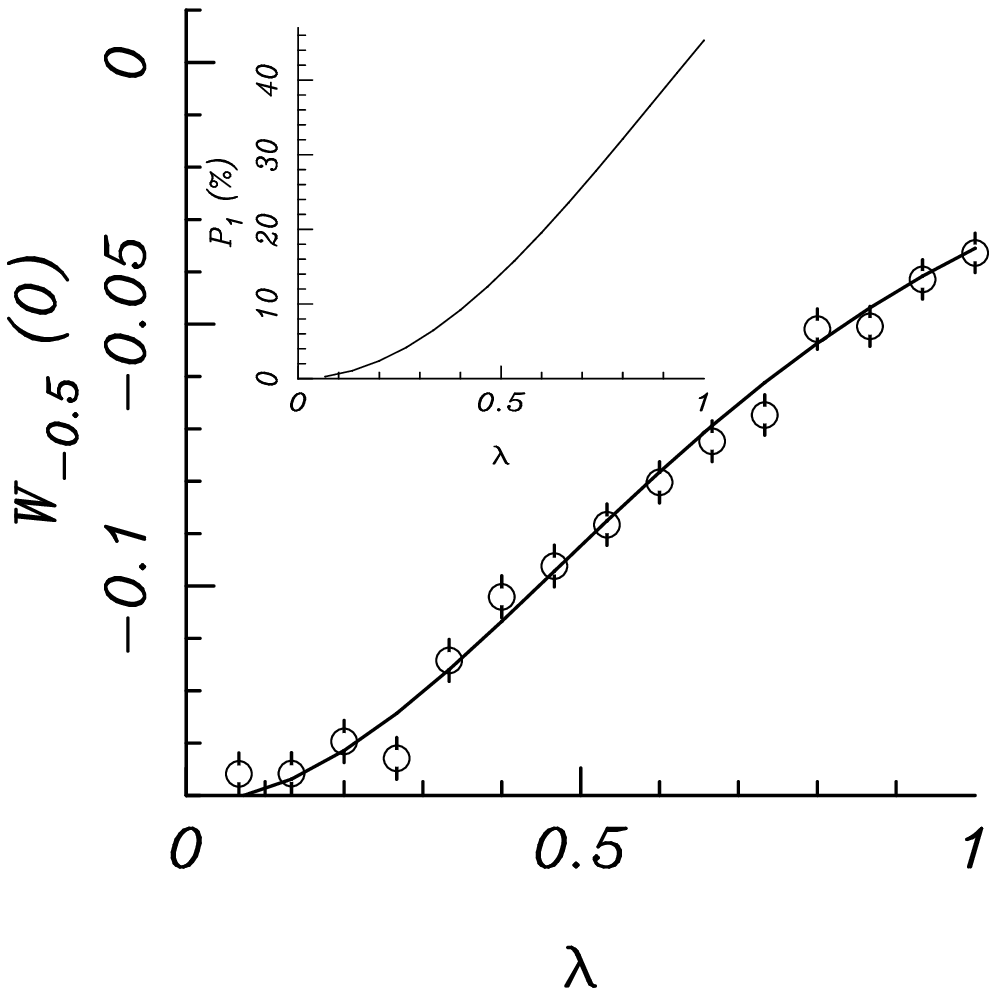,width=4cm} &
\psfig{file=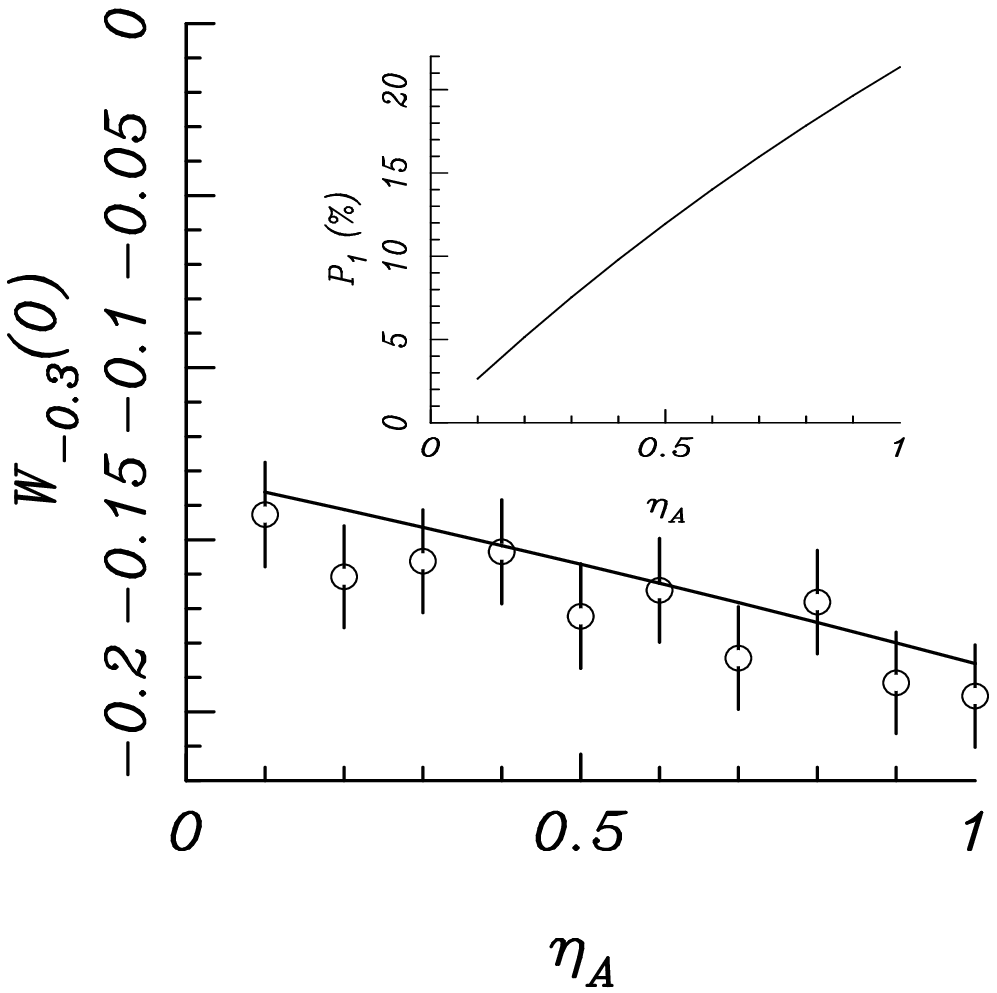,width=4cm}\end{tabular}
\caption{Reconstruction of the Wigner function in the origin of the
phase-space by (Monte Carlo simulated) homodyne tomography (sample of 
$5\times10^4$ data at random phase). Top: 
reconstruction of $W_{-0.5}(0)$ versus gain $\lambda$, for 
$\eta_{\sc a}= 60\%$ and $\eta_{\sc h}=70 \%$. Bottom: reconstruction 
of $W_{-0.3}(0)$ versus the avalanche photodetector quantum efficiency 
$\eta_{\sc a}$, for $\lambda =0.5$ homodyne quantum efficiency 
$\eta_{\sc h}=80 \%$. In both plots the solid line is the theoretical 
value, and the inset shows the detection probability $P_1$.  
\label{f:varEG}}
\end{figure}

\begin{figure}[h]\psfig{file=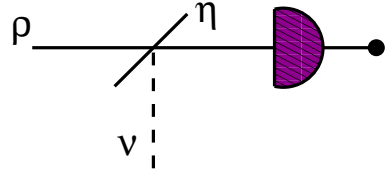,width=5cm}
\caption{Model for a realistic photodetector.\label{f:bsdark}}
\end{figure}

\begin{figure}[h]\hspace{-10pt}\begin{tabular}{lr}
\psfig{file=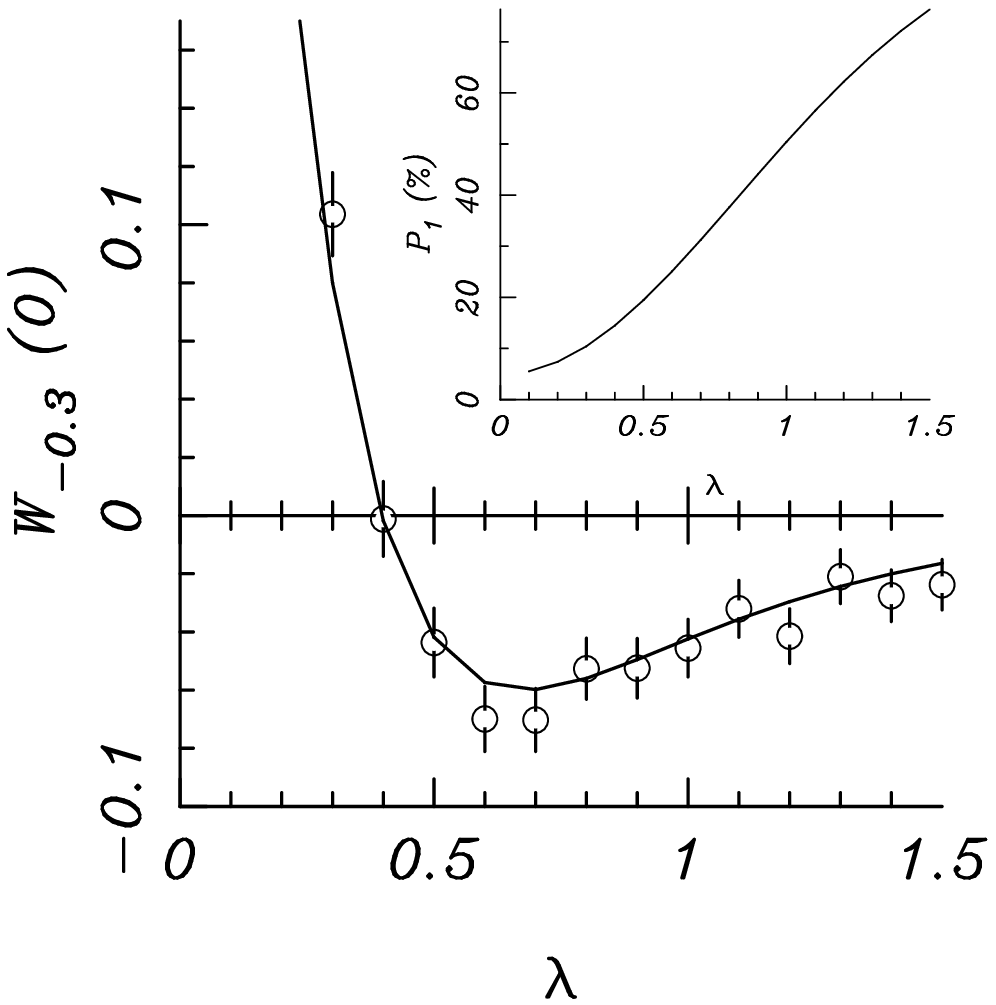,width=4cm} & 
\psfig{file=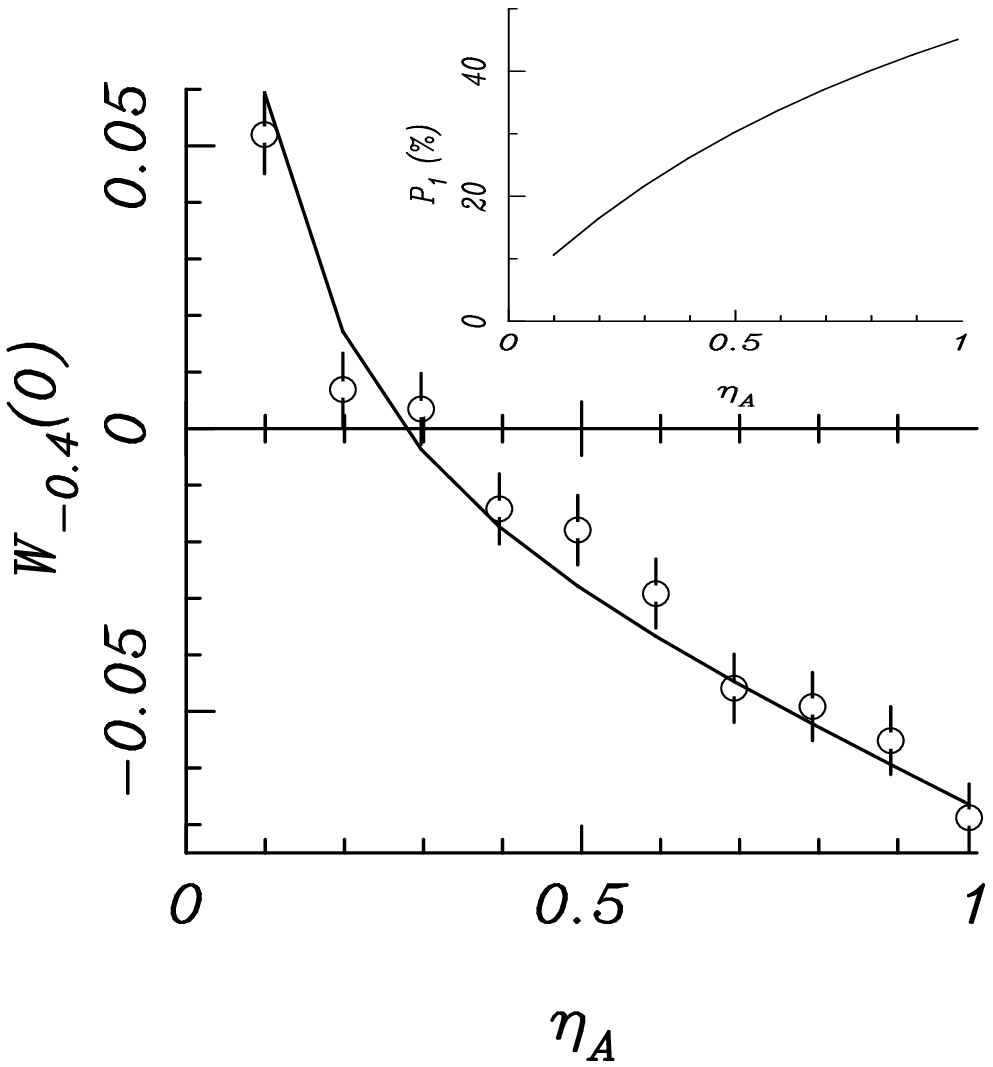,width=4cm}\end{tabular}
\caption{Reconstruction of the Wigner function in the origin of the
phase-space by (Monte Carlo simulated) homodyne tomography (sample of 
$5\times10^4$ data at random phase) Top: reconstruction of $W_{-0.3}(0)$
versus gain $\lambda$, for $\eta_{\sc a}= 70\%$, $\eta_{\sc h}=80 \%$, 
and $N=0.05$. Bottom: reconstruction 
of $W_{-0.4}(0)$ versus quantum efficiency $\eta_{\sc a}$ for $\lambda =0.8$, 
$\eta_{\sc h}=75 \%$, and $N=0.08$. In both plots the solid line is the theoretical 
value, and the inset shows the detection probability $P_1$. 
Increasing $N$ would generally need a larger threshold for $\eta_{\sc a}$
\label{f:darkEG}}
\end{figure}

 
The fact that all the generalized Wigner functions
have negative values may be exploited in order to reveal the
nonclassicality of $\hat\varrho_1$ through quantum homodyne 
tomography. In fact, one has  
\begin{eqnarray}
W_s (0) = \hbox{Tr} \left[ \hat\varrho_1 \: \hat W_s\right] \qquad
\hat W_s = \frac{2}{\pi} \frac{1}{1-s} \left(\frac{s+1}{s-1}
\right)^{a^\dag a} \label{wsop}\;,
\end{eqnarray}
and therefore \cite{bil}
\begin{eqnarray}
W_s (0) = \int dx \: p_{\eta_{\sc h}} (x) \:R_{\eta_{\sc h}}[\hat W_s] (x) 
\label{wstomo}\;,
\end{eqnarray}
where $p_{\eta_{\sc h}} (x)$ is the probability distribution of a random phase
homodyne detection (with quantum efficiency $\eta_{\sc h}$) and 
$R_{\eta_{\sc h}}[\hat W_s] (x)$ is the tomographic kernel for the operator
$\hat W_s$, which is given by \cite{tokyo}
\begin{eqnarray}
R_{\eta_{\sc h}}[\hat W_s] (x) = \frac{2\eta_{\sc h}}{\pi} \frac{\hbox{\large 
$\Phi$}\left(1;1/2; - \frac{2\eta_{\sc h} x^2}{(1-s)\eta_{\sc h} -1}
\right)}{(1-s)\eta_{\sc h} -1}\label{wker}\;,
\end{eqnarray}
where $\Phi(a,b;z)$ is the confluent hypergeometric function.
$R_{\eta_{\sc h}}[\hat W_s] (x)$ is a bounded function for 
$ s < 1 - \eta_{\sc h}^{-1}$ which re\-pre\-sents the maximum 
index of the Wigner function $W_s(0)$ that can be reconstructed by 
homodyne tomography  with efficiency $\eta_{\sc h}$. In Fig. \ref{f:varEG}
we show a typical reconstruction of $W_s(0)$ for different values of $s$, 
$\lambda$ and $\eta_{\sc a}$.
Besides quantum efficiency, i.e. lost photons, the performance of a 
realistic photodetector are also degraded by the presence of dark-count, 
i.e. by "clicks" that do not correspond to any incoming photon. In order 
to take into account both these effects we use the simple scheme depicted 
in Figure \ref{f:bsdark}. A real photodetector is modeled as an ideal 
photodetector (unit quantum efficiency, no dark-count) preceded by a 
beam splitter (of transmissivity equal to the quantum efficiency $\eta$) 
whose second port is in an auxiliary excited state $\hat \nu$, which can 
be a thermal state, or a phase-averaged coherent state, depending on 
the kind of background noise (a thermal or Poissonian). 
If the second port of the beam splitter is the vacuum 
$\hat\nu = |0\rangle\langle 0|$ we have no dark-count and the 
POVM of the photodetector reduces to that of Eq. (\ref{pom}). 
For the second port of the BS excited in a generic mixture $\hat \nu =
\sum_s \nu_{ss} |s\rangle\langle s|$ the POVM for the {\sc on/off}
photodetection is given by 
($\hat\Pi_1\doteq \hat{{\bf I}} -\hat\Pi_0
$) 
\begin{eqnarray}
\hat \Pi_0 = \sum_{n=0}^\infty (1-\eta)^n \sum_{s=0}^\infty \nu_{ss} \: 
\eta^s \: \left(\begin{array}{c} n+s \\ s \end{array}\right) \: 
|n\rangle\langle n| \label{gendark}\;.
\end{eqnarray}
The density matrices of a thermal state and a phase-averaged coherent state 
(with $M$ mean photons) are given by 
\begin{eqnarray}
\hat\nu_{\sc t} &=& \frac{1}{M+1} \sum_s \left(\frac{M}{M+1}\right)^s 
\: |s\rangle\langle s| \\ \hat\nu_{\sc p} &=& e^{-M} \sum_s 
\frac{M^s}{s!}|s\rangle\langle s|  
\label{darkstate}\;. \end{eqnarray}
In order to reproduce a background noise with mean photon number $N$ 
we consider the state $\hat\nu$ with average photon number $M=N/(1-\eta_{\sc a})$. 
In these case we have 
\begin{eqnarray}
\hat\Pi_0^{\sc t} &=& \frac{1}{1+N} \sum_n \left( 1- 
\frac{\eta_{\sc a}}{1+N}\right)^n
\:|n\rangle\langle n| \\  
\hat\Pi_0^{\sc p} &=& e^{-N} \sum_n \Bigg[(1-\eta_{\sc a})^n 
\: L_n (- N \frac{\eta_{\sc a}}{1-\eta_{\sc a}})\Bigg] \:|n\rangle\langle n|
\label{darkpom}\;,
\end{eqnarray}
where ${\sc t}$ and ${\sc p}$ denotes thermal and Poissonian respectively, 
and $L_n(x)$ is the Laguerre polynomial of order $n$. The corresponding 
detection probabilities are given by 
\begin{eqnarray}
P_1^{\sc t} &=& 1- \frac{1 - \xi^2}{[1+N]
(1-\xi^2)+\eta_{\sc a} \xi^2} \\ 
P_1^{\sc p} &=& 1 - \frac{1-\xi^2}{1-\xi^2(1-\eta_{\sc a})}
\exp\left[ - N \frac{1-\xi^2}{1-\xi^2 (1-\eta_{\sc a})}
\right] \label{darkprobs}\;.
\end{eqnarray}
For small $N$ the two models lead 
to the same detection probability at the first order in $N$
$P_1^{\sc t}\simeq P_1^{\sc p}=P_1
+ O[N^2]  $ with
\begin{eqnarray}
P_1 =\frac{\eta_{\sc a} \xi^2}{1-\xi^2 (1-\eta_{\sc a})} +
\frac{(1-\xi^2)^2}{[1-\xi^2 (1-\eta_{\sc a})]^2} N 
\label{asym}\;.
\end{eqnarray}
In the following we will use the Poissonian background, and 
omit the index ${\sc p}$. The conditional output state, after the 
observation of a click, is now given by
\begin{eqnarray}
\hat\varrho_1 = \frac{1}{P_1} (1-\xi^2) &\sum_n & \xi^{2n} \left[
1 - \frac{(1-\eta_{\sc a})^n}{e^N} \right. \nonumber\\  
&\times& \left. L_n (- \frac{N \eta_{\sc a}}{1-\eta_{\sc a}} ) 
\right] \: |n\rangle\langle n| 
\label{darkcond}\;
\end{eqnarray}
and the generalized Wigner function $W_s (0)$ in the origin of the 
phase space is 
\begin{eqnarray}
W_s(0)= \frac{2 (1-\xi^2)}{\pi P_1} && \left\{ 
\frac{1}{(1-s)+\xi^2(1+s)}\right.\label{darkws0} 
\\ &-&\left. \frac{ \exp\left[-N 
\frac{(1+s)+2 \xi^2 (1+s)(1-\eta_{\sc a})}{
(1+s)+\xi^2 (1+s)(1-\eta_{\sc a})}\right]}
{(1-s)+\xi^2 (1-\eta_{\sc a})(1+s)}
\right\}\nonumber \;.
\end{eqnarray}
The Wigner function $W(0)$ for $s=0$ (as well as for $s <0$) is no longer 
negative for all values of parameters. On the other hand, there is 
a large range of values of the 
quantum efficiency $\eta_{\sc a}$ and the gain $\lambda$ giving
negative $W_s(0)$ with $s$ accessible by homodyne tomography 
with realistic values of the homodyne efficiency $\eta_{\sc h}$. 
In other words, the generation and the detection of nonclassicality are 
robust also against the occurrence of dark counts in the avalanche 
conditioning photodetector. 
In Fig. \ref{f:darkEG} we show a typical reconstruction of $W_s(0)$ 
for different values of parameters
In a practical implementation the NOPA consists of a type-II 
phase-matched KTP crystal that is pumped by the second harmonic of a 
Q-switched and mode-locked Nd:YAG laser. Previously, such a NOPA has been 
employed with parametric gains  $>10$ ($|\xi|^2>0.9$), to generate highly 
quantum-correlated twin beams of light at 1064\,nm~\cite{aytur90b}. 
By appropriately choosing the
input quantum state, a similar setup was then used to demonstrate the
production of squeezed-vacuum state with a high degree ($5.8\pm0.2$\,dB) of
squeezing~\cite{kim94b}. In the present context, the twin beams, which
are easily separable because of their orthogonal polarizations resulting
from type-II phase matching, can be separately detected; beam $a$ with a
homodyne detector to verify nonclassicality, and beam
$b$ with an avalanche photodetector. 
The main challenge in the present experiment is the achievement of
high degrees of overlap (mode-matching efficiency) between the down-converted
and the LO modes. Such overlap is non-trivial in pulsed, traveling-wave
experiments owing to the distortion of the down-converted modes that is caused
by the spatio-temporally Gaussian profile of the pump beam. With suitable
choice of LOs, however, $\eta_h>70\%$ has been achieved \cite{aytur92}, an adequate 
value for the present experiment (cf.\ Figs.~\ref{f:varEG} and~\ref{f:darkEG}).
In the measurements on beam $b$, the main challenge will be
the selection of the appropriate mode, which can be performed by 
exploiting quantum-frequency conversion, a process that has been 
previously demonstrated~\cite{huang92}.
\par
In conclusion, we presented a robust experiment to verify the quantum 
nature of light. This goal is achieved by a conditional scheme,
where one of the entangled twin-beam exiting a NOPA is revealed 
by an avalanche photodetector, leaving the other one in a pseudo-thermal 
state with no vacuum component. The nonclassicality, as well as its
verification by homodyne tomography, are robust against amplifier 
gain, detector efficiency and dark counts.  
\par
This work has been cosponsored by MURST (Project "Amplificazione e 
rivelazione di radiazione quantistica") and INFM (PAISS TWIN). The author 
would thank Mauro D'Ariano for many useful discussions and Prem Kumar 
for remarks on the experimental implementation. 

\end{document}